\documentclass[pmlr]{jmlr}

\usepackage{booktabs}
\usepackage{float}   
\usepackage{subcaption}
\usepackage{placeins}%

\jmlrproceedings{Preprint}{Preprint. Submitted to the Conference on Applied Machine Learning for Information Security (CAMLIS) 2026.}
\jmlrpages{}

\title[Secret Scanner Agent]{Secret Scanner Agent: Extracting Secrets and Access Context from Unstructured Documents}

\author{\Name{Zixiao Chen\nametag{\thanks{Corresponding author.}}} \Email{zixiaochen@microsoft.com}\\
   \addr Microsoft, Redmond, WA, USA
   \AND
   \Name{Mariko Wakabayashi} \Email{mwakabayashi@microsoft.com}\\
   \addr Microsoft, Redmond, WA, USA
   \AND
   \Name{Charlotte Siska} \Email{csiska@microsoft.com}\\
   \addr Microsoft, Redmond, WA, USA}

\begin{document}

\maketitle

\begin{abstract}%
Exposed documents such as emails, chat threads, tickets, and incident notes routinely leak credentials, but during incident response a leaked secret is only half the story.
Responders also need to identify the ``door'' the secret opens: the account, tenant, endpoint, database, cloud resource, or other system that the credential could allow an attacker to access.
Traditional secret scanners rely on regular expressions or trained classifiers which work well on well-formatted code, yet they struggle when a credential is fragmented, reformatted, or far from the resource it unlocks, and they report the secret string without naming what it opens.
We present Secret Scanner Agent (SSA), a multi-agent large-language-model system that extracts both the secret and its associated door, together with supporting evidence, from unstructured exposed documents.
SSA pairs a detection agent that favors recall with a review agent that filters false positives and recovers missing context.
Because real credential data is sensitive, we evaluate SSA on synthetic benchmarks we generated that span 23 secret types and multiple document formats, scored with a three-step pipeline of programmatic matching, an LLM judge, and human review.
Across six models, multi-agent SSA improves extraction precision over a single-agent variant, with the largest gains on door extraction, by up to 16 percentage points.
SSA matches a regular-expression scanner's precision while more than tripling its recall, and against thirteen security analysts it is more precise, recovers nearly twice as many secret--door pairs, and runs five to seventeen times faster.
By returning the secret, its door, and supporting evidence in one result, SSA turns credential detection into an actionable finding for triage and remediation.
\end{abstract}


\section{Introduction}
\label{sec:introduction}

Security teams need to answer three questions during a breach or data exposure: (1) did the exposed material contain secrets, (2) what systems or resources do those secrets unlock, and (3) what action should responders take? Secrets are credentials or authentication materials that grant access to systems, services, data, or infrastructure. They can include passwords, API keys, access tokens, private keys, cloud account keys, database connection strings, certificates, and other values used by users, applications, or services to prove identity \citep{owasp_secrets_cheatsheet, ibm_secrets_definition}. When attackers find exposed secrets, they may impersonate users or services, access sensitive data, move laterally across systems, or maintain persistence inside an environment. Recent incidents show that this risk extends beyond source code and into exposed documents, support artifacts, emails, tickets, and logs.

\begin{figure}[ht!]
\floatconts
  {fig:secret_exposure_and_triage}
  {\caption{Secret exposure and response workflow. Exposed secrets in
   unstructured content create both access risk and triage burden.}}
  {%
    \subfigure[Secrets exposed in unstructured content can unlock downstream
      systems.]{%
      \label{fig:exposed_secrets_threatsa}%
      \includegraphics[width=0.95\textwidth]{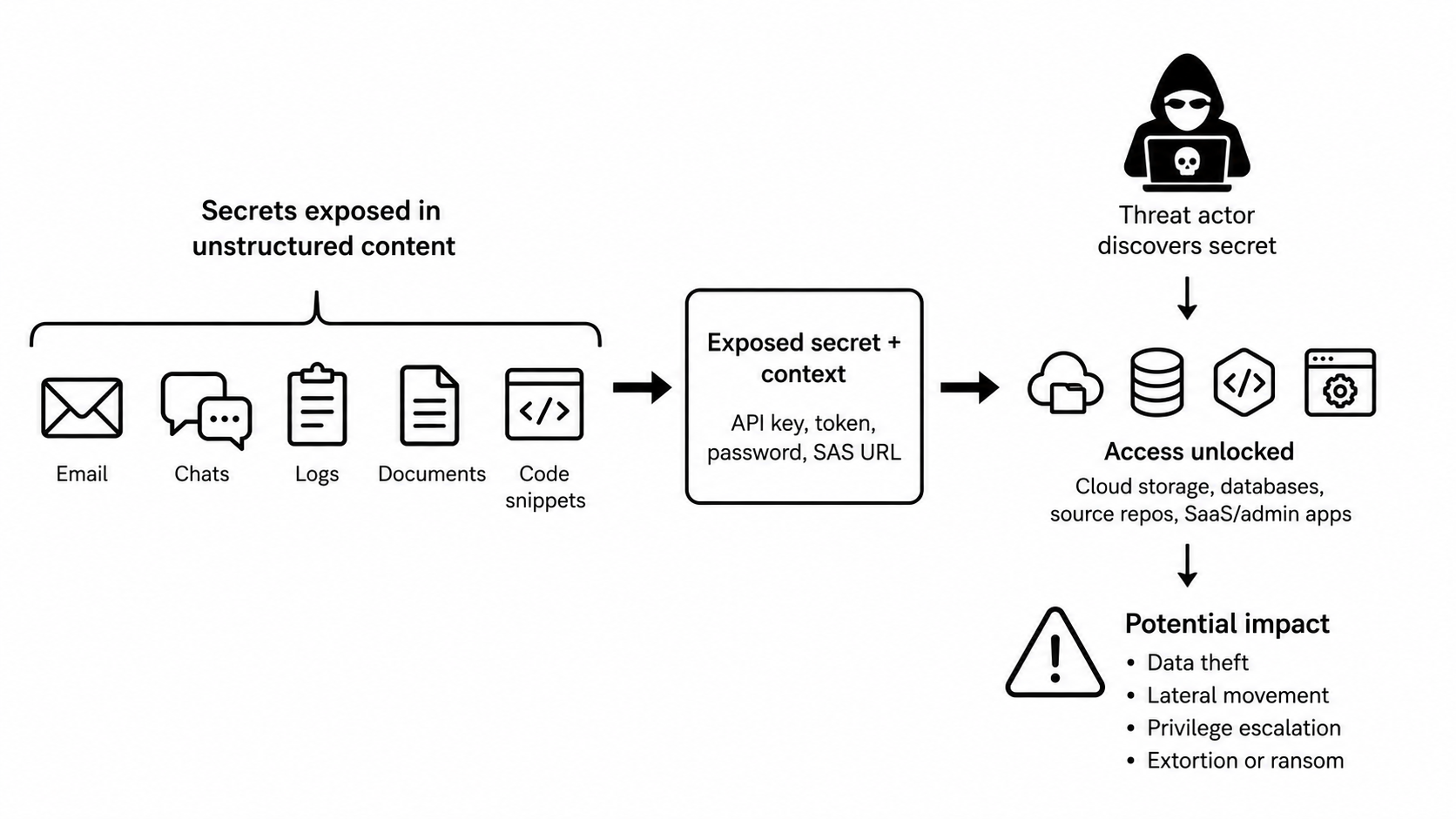}}\\[0.5cm]
    \subfigure[Responders must validate, revoke, and investigate exposed
      credentials.]{%
      \label{fig:exposed_secrets_triageb}%
      \includegraphics[width=0.95\textwidth]{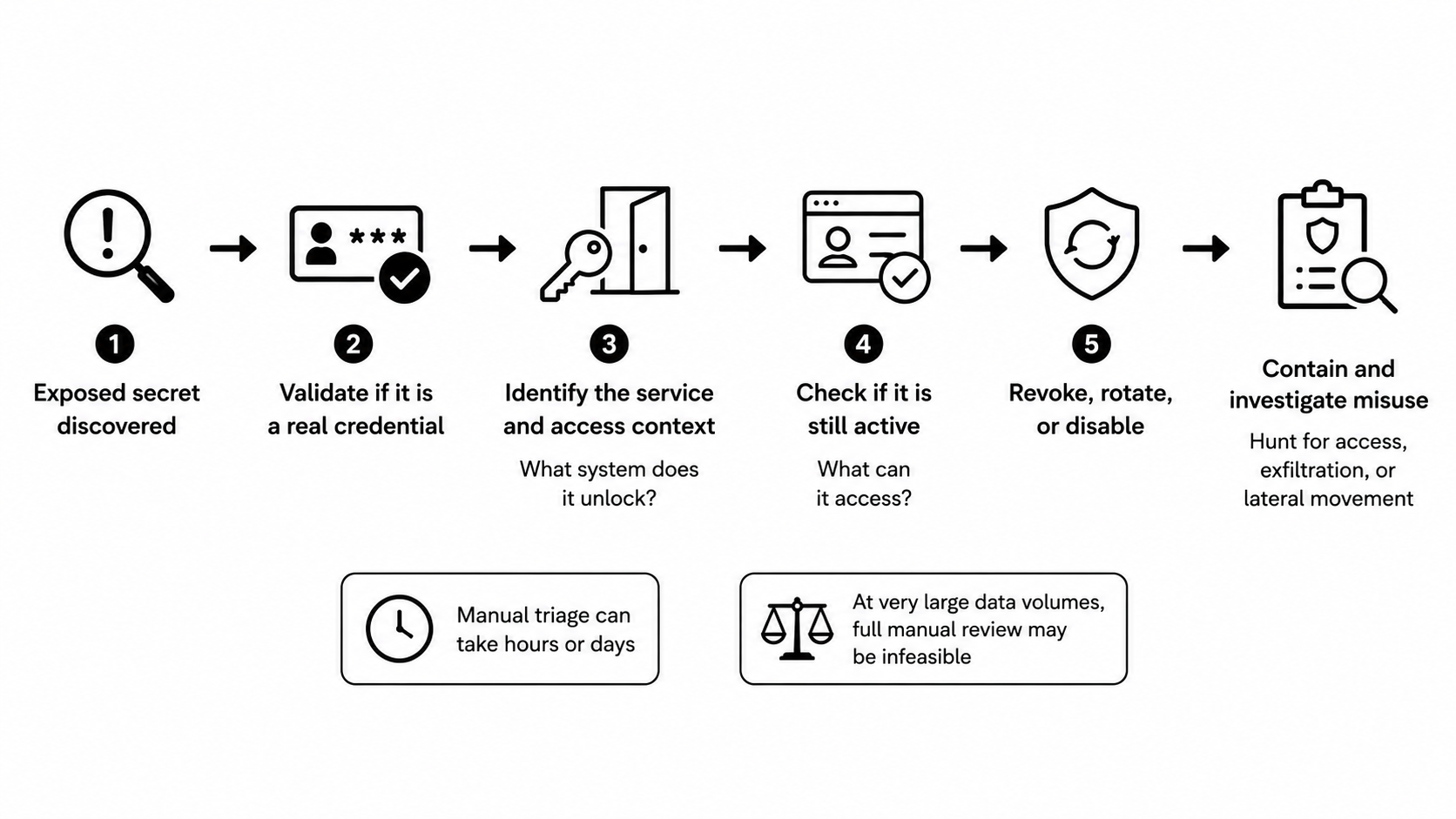}}
  }
\end{figure}

The 2023 Okta support case management incident illustrates this risk. Okta disclosed that a threat actor accessed files uploaded by customers as part of recent support cases, including HTTP Archive (HAR) files that can contain sensitive browser data such as cookies and session tokens \citep{okta_support_case_management_2023}. Cloudflare later reported that, in its case, the threat actor used one access token and three service account credentials taken during the Okta compromise, which Cloudflare had not rotated, to access parts of Cloudflare’s internal Atlassian environment \citep{cloudflare_thanksgiving_security_incident_2023}. A similar risk appeared in the Microsoft Midnight Blizzard incident, where Microsoft reported that information initially exfiltrated from corporate email systems was later used to gain, or attempt to gain, unauthorized access to source code repositories and internal systems \citep{microsoft_midnight_blizzard_2024_jan,microsoft_midnight_blizzard_2024_mar}. These incidents share a common pattern: exposed documents can contain both credentials and the surrounding context needed to use them. As illustrated in Figure~\ref{fig:exposed_secrets_threatsa}, attackers can mine support files, emails, tickets, and troubleshooting artifacts for credentials, resource names, internal systems, endpoints, tenants, and other context that may support follow-on access.

Given this risk, existing secret-finding tools provide an important first line of defense. Tools such as GitHub Secret Scanning, TruffleHog, Gitleaks, Yelp's detect-secrets, and GitGuardian commonly combine regular expressions, provider-specific patterns, entropy checks, keyword heuristics, contextual signals, validation workflows, and in some cases machine learning to identify exposed credentials \citep{github_secret_scanning,trufflehog_github,gitleaks_github,yelp_detect_secrets,gitguardian_secret_detection}. These approaches work well when a secret appears in a recognizable format, such as a known token prefix, a private key block, a cloud access key, or a random-looking string near words such as \texttt{password}, \texttt{token}, \texttt{secret}, or \texttt{api\_key}. They are fast, scalable, and easy to integrate into developer and security workflows.

However, exposed documents are harder to analyze because they often break the format, keyword, and context assumptions that many secret-finding tools rely on. Emails, tickets, chat threads, support conversations, incident notes, internal documents, pasted logs, screenshots, quoted replies, and follow-up corrections often mix natural language with technical artifacts \citep{valadon2024hardcoded_secrets}. A credential may appear in an incomplete format, span multiple messages, or sit far away from the resource it unlocks. A document may contain enough evidence for a human analyst to understand the exposure, but not in a structure that traditional scanners can reliably parse. As a result, responders can face both missed findings and noisy alerts.

Even when a secret is detected, responders still need to answer a second question: what does the secret unlock? During incident response, teams need to identify the associated ``door'': the account, tenant, service, endpoint, database host, storage bucket, cloud resource, or system that the secret may unlock \citep{owasp_top10_2021}. \appendixref{app:example} shows a concrete synthetic example of this: an unstructured request to rotate a storage-account key, together with the secret, door, and supporting evidence extracted from it. A secret by itself tells responders that something sensitive may have leaked. A secret and door pair tells responders what may be exposed and where to act. Validation can help determine whether a credential is active, but it often depends on knowing both the secret type and the associated resource. As shown in Figure~\ref{fig:exposed_secrets_triageb}, responders must then validate the finding, identify what the credential unlocks, determine whether it is still active, revoke or rotate it, and investigate possible misuse.

This workflow can be slow and difficult to perform manually, especially at large scale. We therefore explore how large language models can help responders extract actionable secret and door pairs from unstructured exposed documents. We began with a single-agent approach to identify secrets and related doors in messy documents. However, it could miss important context, extract incomplete information, or fail to connect a credential with the correct resource. These failure modes motivated a multi-agent workflow that separates candidate extraction, review, and verification, making secret and door extraction more reliable for incident-response scenarios.

Evaluation presents another challenge. Real customer data is difficult to access and share, especially for sensitive security problems involving credentials. To evaluate earlier and iterate safely, we used realistic synthetic documents generated with a self-reflection framework similar to DataGen, extended with an additional agent that injected secrets and doors \citep{huang_ICLR2025_a01e69aa}. This allowed us to test extraction across messy document formats, study failure modes, and build confidence before exploring product integration.

Our results show that agentic secret finding can help responders move from detecting exposed credentials to understanding what those credentials may unlock. By recovering the secret, the associated door, and supporting evidence from artifacts such as emails, tickets, chat threads, and incident notes, the approach can provide more actionable context for validation, triage, and remediation.

This work makes five contributions to secret detection in unstructured exposed documents:

\begin{itemize}
\item \textbf{SSA system design:} We introduce Secret Scanner Agent (SSA), an automated and scalable approach for extracting actionable secret-exposure context from unstructured content such as emails, tickets, chat threads, and incident notes.

\item \textbf{Secret and door problem formulation:} We frame secret finding as more than credential detection. SSA extracts both the exposed secret and the associated door, such as an account, tenant, endpoint, database host, storage bucket, cloud resource, or internal service that the secret may unlock.

\item \textbf{Synthetic benchmark pipeline:} We build a synthetic data generation pipeline for creating realistic emails that contain secrets, doors, and surrounding context. This enables safer evaluation when real customer data is sensitive, rare, or difficult to share.

\item \textbf{Evaluation framework:} We propose an evaluation framework for agentic secret finding that measures both detection and extraction quality: whether unstructured content contains a secret and/or door, and whether the method extracts the exact secret and/or door.

\item \textbf{Empirical evaluation:} We compare SSA against baseline secret-finding tools and measure the benefit of the multi-agent workflow.
\end{itemize}

\section{Related Work}
\label{sec:related-work}

\subsection{Secret finding}
\label{sec:rw-secret-finding}
Prior work in secret finding has largely focused on identifying exposed credentials in source code, configuration files, commits, logs, and developer workflows. Common tools such as GitHub Secret Scanning, TruffleHog, Gitleaks, Yelp's detect-secrets, and GitGuardian combine pattern matching, entropy checks, keyword heuristics, contextual signals, validation workflows, and, in some cases, machine learning \citep{github_secret_scanning,trufflehog_github,gitleaks_github,yelp_detect_secrets,gitguardian_secret_detection}.

Pattern-based approaches work well when credentials appear in recognizable formats \citep{ahmed2026secretleakdetectionsoftware}. Provider-specific detectors match known prefixes, lengths, delimiters, or character sets, while generic detectors look for suspicious values near terms such as \texttt{password}, \texttt{token}, \texttt{secret}, \texttt{credential}, \texttt{api\_key}, or \texttt{connection\_string}. Entropy checks add another signal by flagging random-looking strings that may represent keys or tokens. These methods are fast, scalable, explainable, and easy to integrate into developer and security workflows.

However, these approaches still face two recurring challenges: noisy findings and missed credentials. False positives occur when identifiers, UUIDs, hashes, placeholders, test values, or sample tokens resemble secrets but are not usable \citep{Basak2023ACS}. Missed credentials occur when a secret doesn't match an expected format, appears incomplete, or is split across lines, messages, or sections. These problems become harder in unstructured exposed documents, where natural language and technical artifacts are often interleaved across emails, tickets, chat threads, pasted logs, screenshots, and quoted replies. The evidence may be clear to a human analyst, but not organized in a form that a scanner can reliably parse.

Validation workflows help reduce uncertainty by checking whether a detected credential is active or usable. GitHub Secret Scanning supports validity checks for supported patterns, and TruffleHog describes verification as checking whether a credential can authenticate to the issuing service \citep{github_validity_checks,trufflehog_verification}. However, validation and remediation often require more than the secret value alone. Some checks depend on knowing the credential type or issuing provider, while impact assessment and remediation often require the associated resource, account, endpoint, or service. This makes secret and door pairing important: without the door, a detected secret may still require manual investigation before responders know what is exposed or where to act.

Machine learning approaches use additional context to improve detection and prioritization. Nightfall describes AI-powered secret scanning across developer tools and SaaS platforms such as GitHub, Slack, and Jira \citep{nightfall_snyk_ai_scanning}, while GitGuardian describes machine learning features that remove false positives, enrich generic secret findings, and score incident risk \citep{gitguardian_ml,gitguardian_risk_score}. These systems reflect a shift from string matching alone toward context-aware secret detection. However, traditional machine learning approaches often depend on clean training data, well-formed examples, and representative labels \citep{bay2024machinelearningvsdeep, peng2025interpretingcursedimensionalitydistance, mikolajczykbarela2023surveybiasmachinelearning}. In security settings, such data is difficult to collect because real customer data is sensitive, rare, and hard to share. They may also struggle with fragmented credentials, partial redactions, informal developer conversations, and messy email threads, or produce scores without analyst-friendly supporting evidence.

\subsection{Large language models for secret finding}
\label{sec:rw-llm}
Large language models offer a complementary direction for analyzing unstructured exposed documents. They can read natural language, logs, developer discussion, incident notes, and technical artifacts together. They can use surrounding evidence to identify candidate credentials even when it doesn't appear in a standard format, which helps reduce the false positives that regex- and entropy-based detectors produce \citep{rahman2025secretbreachsourcecode, biringa2025hardcodedcredentialsllm, ahmed2026secretleakdetectionsoftware}. Recent studies apply this approach across source code, mobile apps, software issue reports, and even preprint archives \citep{alecci2025evaluatingllmsecretsandroid, dubniczky2025latexposed}. They can also point to supporting context, such as nearby resource names, URLs, account identifiers, service names, or operational instructions. This evidence doesn't replace validation, but it can give analysts more context than a score alone.

LLMs can also help connect credentials to related doors, such as storage accounts, cloud accounts, database hosts, API endpoints, customer tenants, or internal service names, especially when evidence is scattered across human communication rather than stored in a predictable structure. Our work focuses on this operational gap: extracting the secret, associated door, and supporting evidence from unstructured exposed documents so responders can validate, triage, and remediate more quickly.

\subsection{Multi-agent frameworks}
\label{sec:rw-multi-agent}

Multi-agent approaches coordinate several large-language-model agents to solve tasks that a single prompt handles poorly. AutoGen frames this coordination as a conversation among configurable agents that exchange messages, call tools, and optionally involve a human \citep{wu_autogen_2023}. Other frameworks assign agents distinct roles: CAMEL uses role-playing agents that cooperate toward a shared goal \citep{li_camel_2023}, while MetaGPT and ChatDev organize role-specialized agents around software-engineering workflows \citep{hong_metagpt_2024, qian_chatdev_2024}. Across these systems, dividing work among specialized agents can improve reliability on complex, multi-step tasks.

Prior work also shows the value of evidence gathering and critique loops. ReAct interleaves reasoning with actions so an agent can gather evidence before committing to an answer \citep{yao_react_2023}. Reflexion and Self-Refine use critique and revision to improve outputs without additional training \citep{shinn_reflexion_2023, madaan_selfrefine_2023}, while multi-agent debate lets separate agents propose and challenge answers \citep{du_debate_2023}. SSA brings this detect-then-critique pattern to secret finding: a detection agent favors recall and surfaces candidate secrets and doors, and a review agent checks each candidate against the source, removes false positives, and recovers missing context. To our knowledge, prior work hasn't applied this detect-then-critique multi-agent pattern to extracting secrets, associated doors, and supporting evidence from unstructured exposed documents.

\section{Data}
\label{sec:data}

Availability of appropriate datasets is an ongoing issue when evaluating secret-finding methods.
Scarce available data is an ongoing issue in security given the highly sensitive nature of real-world samples.
Researchers have explored generating synthetic datasets with language models through in-context learning, fine-tuning, and introducing perturbations to solve this same issue across various disciplines \citep{abaskohi_utnlp_2022, liu_wanli_2022, lupidi_source2synth_2024, huang_ICLR2025_a01e69aa}.
Here, we have generated our own samples using standard synthetic data generation techniques.

\subsection{Seed Data}
\label{sec:data-support-emails}
The data we use as a seed for synthetic data generation are open source scrubbed customer support emails.\footnote{\url{https://github.com/karolzak/support-tickets-classification}}
The raw data has 20K samples.
Samples are clustered using cosine similarity of ada-2 text embeddings,\footnote{\url{https://openai.com/index/new-and-improved-embedding-model/}} which yields about 456 clusters. Of these clusters, one sample is randomly chosen.
We generate three samples for each cluster, resulting in 1350 emails.

\begin{figure}[htbp]
\floatconts
  {fig:self_reflection}
  {\caption{Self-Reflection Framework}}
  {\includegraphics[width=0.3\textwidth]{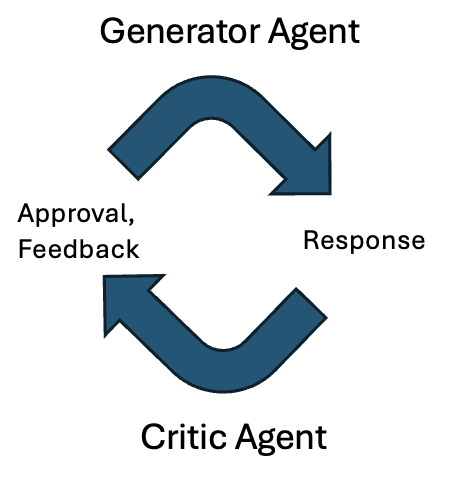}}
\end{figure}

\subsection{Synthetic Data Generation}
\label{sec:data-agentic}

To generate synthetic data, we use a self-reflection framework and in-context learning similar to the pipeline from DataGen in \citet{huang_ICLR2025_a01e69aa}.
The data generation pipeline is built using AutoGen to facilitate self-reflection where agents communicate in a loop \citep{wu_autogen_2023}.
Self-reflection consists of two agents: the generator agent that creates the response and the critic agent that reviews the response as presented in \figureref{fig:self_reflection}.
The agents loop through multiple cycles until the response has been approved by the critic agent.

In this work, the response is a synthetic email generated with in-context learning with seed customer support email samples (\figureref{fig:data_pipeline}).
The critic agent judges the synthetic email based on (1) how organic the email is, (2) how similar the email is to the provided few-shot samples and (3) how distinct samples are from the original samples.

\begin{figure}[htbp]
\floatconts
  {fig:data_pipeline}
  {\caption{Synthetic Email with Secrets Pipeline}}
  {\includegraphics[width=0.8\textwidth]{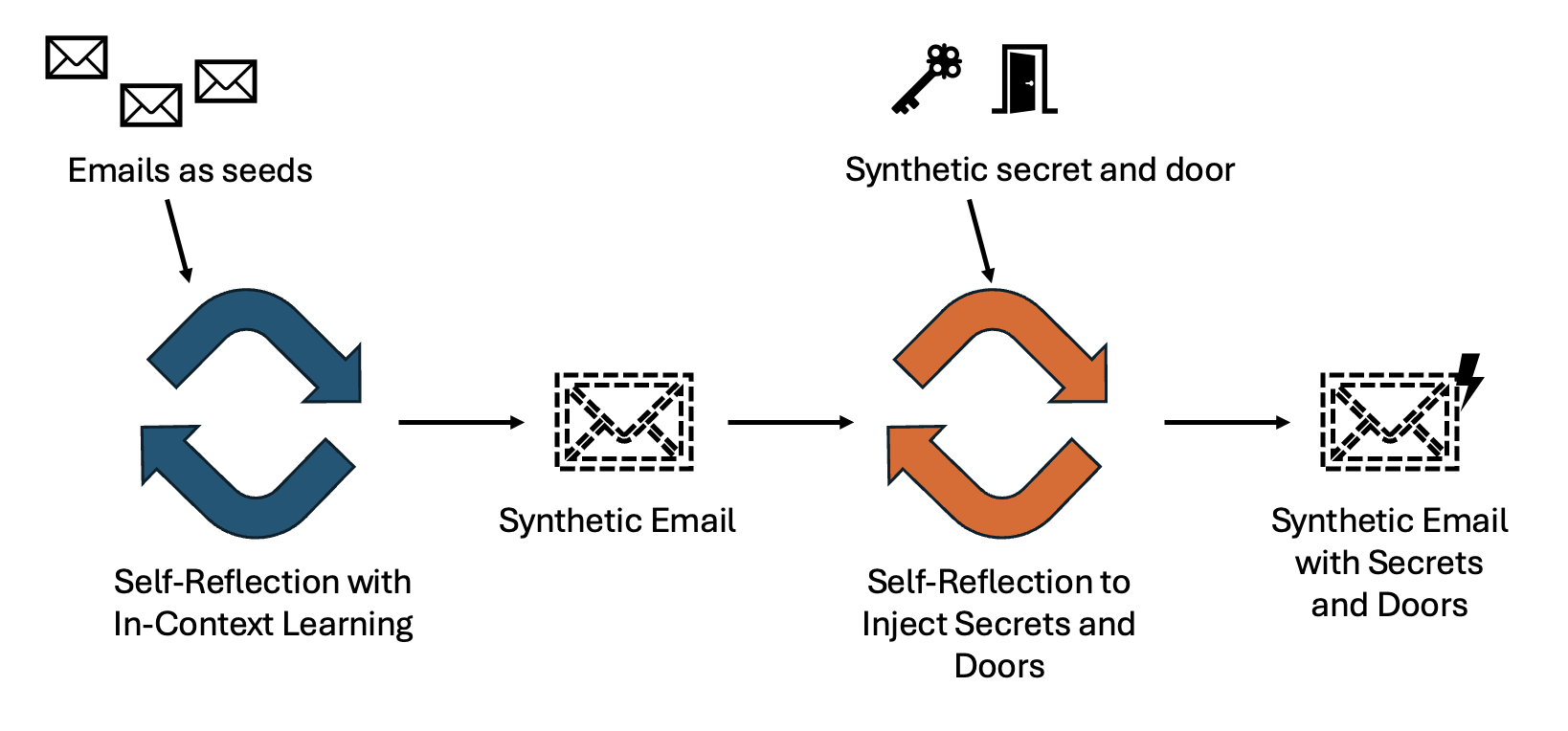}}
\end{figure}

To generate fake secrets and doors, we use regular expressions to find secrets by Trufflehog and reverse them to generate strings \citep{trufflehog_github}.
These strings are in the same format as dictated by the regular expressions, and therefore mimic secrets and doors in the real-world without the threat of secrets being usable.
Another self-reflection loop is used to inject the secret, where the critic agent judges the injection based on how obvious it is, as shown in \figureref{fig:data_pipeline}.

All following benchmarks are generated using this pipeline.
For each document, the content body, embedded secrets, and associated doors are synthetic; no secret corresponds to a real account or live resource. Each secret and door is sampled to match the regular expression pattern of a genuine credential type and its associated resource.

\subsection{Benchmark datasets}
\label{sec:benchmark-datasets}

We evaluate SSA on three complementary benchmarks. The first is our main multi-type benchmark: 356 synthetic emails, including 15 positives and 341 benign examples, covering 23 major secret types. The second benchmark isolates one credential class, Azure storage account keys, across 204 emails, including 19 positives and 185 benign examples. The third benchmark supports the human-agent review study and broadens the document formats beyond email. It contains 45 documents spanning emails, chat transcripts, Word documents, and notes.

\paragraph{Multi-type benchmark.}
The main benchmark contains 356 synthetic emails. 15 emails contain at least one secret-door pair and 341 are benign. Each positive email contains multiple secret-door pairs, for a total of 36 secret instances across the benchmark, and the positive set collectively covers 23 major secret types, each paired with the corresponding kind of door. The full list of secret types is provided in the appendix.

This benchmark tests two capabilities at once. First, SSA must identify which emails contain exposed credentials. Second, for positive emails, SSA must extract all secret-door pairs from documents that may contain several credential types in the same email.

\paragraph{Single-type benchmark.}
The single-type benchmark contains 204 synthetic emails focused exclusively on Azure storage account keys. It includes 19 positive examples and 185 benign examples. Every positive email contains at least two secret-door pairs.

This benchmark isolates SSA's performance on one common, high-risk credential type. Contrasting it with the multi-type benchmark isolates the effect of credential diversity on performance.

\paragraph{Human-review dataset.}
The human-agent review dataset is a separate synthetic benchmark used to compare SSA with human security experts on the same inputs. The dataset contains 45 documents: 33 emails, 4 chat transcripts, 4 Word documents, and 4 notes. All secrets in this benchmark are synthetic Azure storage account keys, and all doors are synthetic Azure storage account URLs for the corresponding storage accounts. Of the 45 documents, 23 are positive and 22 are benign. Positive emails typically contain two or three secret-door pairs.

\paragraph{Challenging cases.}
To evaluate SSA on cases that require more than pattern matching, we include challenging examples that introduce ambiguity, missing context, or competing evidence. These examples cover the following categories:
\begin{itemize}
\item Fragmented key: the secret is split across multiple spans and must be reconstructed.
\item Incomplete key: the apparent secret is truncated or missing required characters.
\item Incomplete door: the associated resource is truncated or missing required components.
\item Distractor key: an irrelevant credential-like string appears alongside the true secret.
\item Distractor door: an irrelevant resource appears alongside the true door.
\item Shared key: the same secret is associated with multiple doors.
\item Stale or superseded key: an initially shared key is later corrected or replaced.
\end{itemize}
This tests whether SSA can use conversational context rather than extracting credentials purely by surface form.

\section{Methods}
\label{sec:methods}

\subsection{SSA: Single-Agent and Multi-Agent Variants}
\label{sec:methods-asf}

SSA has two variants, shown in \figureref{fig:asf-architecture}: a single-prompt baseline and a multi-agent workflow. Both variants read an unstructured document and return any extracted secrets, associated doors, and supporting evidence. They differ in whether the initial detection output is returned directly or reviewed by a second agent before producing the final result.

We first implemented a single-prompt, single-agent baseline to measure whether one LLM call could solve the task and to establish a direct comparison point for the multi-agent workflow. This also helped us avoid adding orchestration complexity unless it produced measurable value. The baseline handled straightforward examples, but failure analysis showed recurring issues: it sometimes extracted incomplete credentials, missed doors that appeared far from the secret, included credential-like distractors, or paired a secret with the wrong resource. These failure modes motivated the multi-agent variant. Rather than asking one model call to detect candidates, infer doors, filter false positives, and produce final structured output in a single pass, multi-agent SSA separates candidate extraction from review. The review agent checks the detector output against the source document, removes unsupported candidates, and recovers missing context when possible.

\begin{figure}[t]
\floatconts
  {fig:asf-architecture}
  {\caption{The two SSA variants. Both take a document and output its secrets and
   the door each one unlocks. The single-agent workflow runs one secret detector
   agent, while the multi-agent workflow adds a verification agent that reviews
   the detector's output before the result is returned.}}
  {\includegraphics[width=\linewidth]{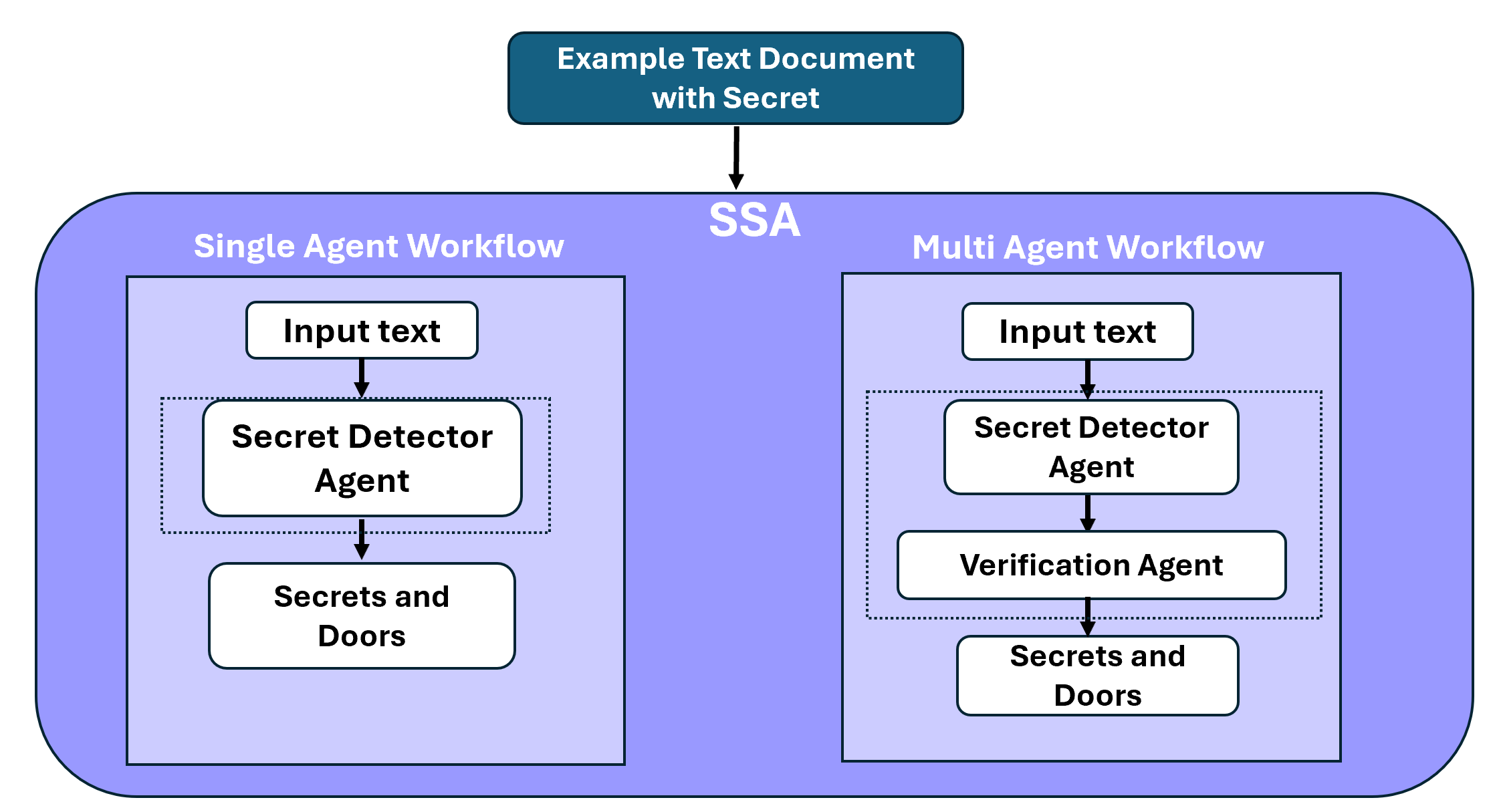}}
\end{figure}

The single-agent variant performs the full task in one model call: it reads the document, identifies candidate secrets, infers associated doors, and writes the structured result. This makes it lower latency and cheaper to run, but it also means detection, filtering, door inference, and output formatting all happen in a single pass without an explicit review step.

The multi-agent variant addresses these failure modes by separating recall-oriented candidate detection from evidence-oriented review. Multi-agent SSA uses two agents. The first agent performs recall-oriented detection: it scans the input document for candidate credentials and related contextual signals, including values that may be incomplete, split across messages, or written in an unusual format.

The second agent reviews the detector output against the original document. It checks whether each candidate is supported by source evidence, filters values that only resemble secrets, and verifies whether the associated door is grounded in the document. This review step helps remove false positives such as placeholders, test values, UUIDs, hashes, and unrelated random-looking strings. It can also recover missing context when the detector identifies a secret but misses the resource, account, endpoint, or service the secret may unlock.

This two-agent design makes SSA more robust for incident-response scenarios. The detection agent focuses on broad candidate discovery, while the review agent focuses on correctness, completeness, and evidence. As a result, multi-agent SSA returns not only a secret-looking value, but also the associated door and supporting context. This gives responders a more actionable result for validation, triage, and remediation because they can see both what credential may have leaked and what system, account, or resource it may unlock.

\subsection{Language models}
\label{sec:methods-models}

We run SSA with six model configurations. Five come from a single provider
family \citep{openai_gpt41_2025, openai_gpt5_2025}: \texttt{gpt-4.1-mini}, \texttt{gpt-5.4-mini} at
medium and high reasoning effort, and \texttt{gpt-5.4-nano} at medium and high
reasoning effort.  The sixth, \texttt{glm-5.2}, is a larger open-weight model from
a different family \citep{glm2024}. The five proprietary configurations span two
model generations, two model sizes, and two reasoning-effort settings, letting us
observe how secret and door extraction responds to changes in model capability
and cost while the rest of the pipeline stays fixed; \texttt{glm-5.2} then tests
whether those observations extend beyond a single vendor and size class.

We keep this core set within one model family rather than surveying every
available frontier model. The aim is not to identify the strongest large language
model for this task, but to test whether agentic secret finding improves on
established secret-finding tools and to learn how much model capability SSA needs
to reach that bar. Holding the family fixed for this comparison removes confounds
from differing tokenizers, prompt formats, and provider behavior, so the
within-family differences we report track model size and reasoning effort rather
than vendor-specific quirks.

We also choose smaller, cheaper, and faster models by design. A system that scans
exposed documents at scale may process millions of artifacts, where per-document
cost and latency decide whether it is deployable in production rather than only in
an offline study. The mini and nano models we evaluate are inexpensive and fast
enough to sit inside everyday response workflows. Demonstrating that SSA performs
well with these models, rather than only with the largest available ones, supports
our central claim that agentic secret finding can be a practical and broadly
available layer of defense.

We include \texttt{glm-5.2} for two reasons. The first is openness: every other
model in our study is proprietary and reachable only through a vendor API, so an
open-weight model lets us ask whether SSA's behavior depends on a single provider
or carries over to models a team could host itself. The second is scale. Our main
set deliberately leans small, and \texttt{glm-5.2} lets us look in the other
direction, at a high-capacity model from an unrelated family, and check whether
the same detection and review prompts still hold without any per-model tuning. If
SSA performs well across this range, from small proprietary models to a large
open-weight one, its benefits are more likely to reflect the agentic workflow
itself than the characteristics of one model family.

\subsection{Baselines: traditional secret-finding tools}
\label{sec:methods-baselines}

We compare SSA against two established secret scanners, TruffleHog and Nightfall
\citep{trufflehog_github,nightfall_snyk_ai_scanning}. All three share the same goal
of finding exposed secrets in text, but they detect them in different ways and
return different output.

TruffleHog is an open-source scanner that detects credentials with a large library
of provider-specific regular-expression detectors, classifies the matched secret by
type, and can confirm a finding by authenticating the credential against its
issuing API \citep{trufflehog_verification}. Its regex detection is deterministic,
so the same document always produces the same result, which is why we ran it once.
We scanned each document with the filesystem subcommand and the default detector
set. Because our secrets are synthetic and not live, we use TruffleHog's pattern
detection rather than its verification step, which cannot confirm a credential that
doesn't authenticate against a real service.

Nightfall is a commercial data-loss-prevention platform that classifies sensitive
content, including secrets and credentials, with trained machine-learning detectors
rather than fixed patterns. We scanned each document through Nightfall's text
scanning API using a single detection rule, and set each detector's minimum
confidence to \texttt{possible}, a permissive threshold that doesn't understate
recall, and averaged results over five runs. Documents larger than the API request
size limit were split into smaller chunks before scanning.

SSA differs from both on two axes. It is large-language-model based and agentic
rather than pattern- or classifier-based, so it reasons over the surrounding text
instead of matching a fixed form, which helps on fragmented, misformatted, or
context-dependent secrets that defeat a regex. It also doesn't stop at the secret:
SSA detects and extracts the associated door, naming the resource the credential
unlocks and giving an analyst the risk and a place to act. Neither TruffleHog nor
Nightfall identifies doors.

These approaches trade off in predictable ways. TruffleHog and Nightfall are
lightweight and fast, close to instant per document, and TruffleHog's regex
matching is fully deterministic, which makes them cheap to run at scale. SSA uses
more time and compute per document in exchange for context-aware detection and the
secret--door pairing. We evaluate all three on the multi-type secret dataset, and
\sectionref{sec:results-baselines} reports how the tradeoff plays out on precision
and recall.

\subsection{Human review}
\label{sec:methods-human-review}

We ran a human-review study to compare SSA against expert effort on the same task,
using the human-review benchmark of 45 documents. Thirteen security analysts
volunteered to take part, with an average of 9.88 years of experience in the
cybersecurity domain, including hands-on work with credentials in operational
settings. We asked each analyst to read every document and record every secret
together with its associated door, producing a final list of secret--door pairs.
Each analyst reviewed all 45 documents independently without any secret scanning
tool or AI assistance.

We gave multi-agent SSA the identical task on the same documents and scored both
against the same ground truth, so the only difference between the two sides is who
or what produced the findings. We recorded the time each analyst spent on the full
set and converted it to a per-document latency by dividing by the document count,
which makes it comparable to SSA's per-document runtime.

We score both sides the same way. Because an analyst produces a final list of
secret--door pairs rather than a staged detection-then-extraction pipeline, we
compare at the level of extracted pairs. A pair counts as correct only when both
the secret and its associated door match the ground truth, under the same matching
rules we apply to SSA. \sectionref{sec:results-human} reports precision, recall,
and latency for both sides.

\subsection{Evaluation pipeline}
\label{sec:methods-eval}

We evaluate four subtasks under precision and recall: secret detection, door
detection, secret extraction, and door extraction. Detection asks whether a secret
or its door is identified, and extraction asks whether its exact value is recovered.
Because large language model outputs are nondeterministic, we run every
configuration five times and report the mean with a 95\% confidence interval.

Scoring extraction is complicated by the free-form nature of model output. The model
frequently returns the correct credential wrapped in extra text or reformatted, so a
naive exact-string comparison would mark a correct answer as wrong. A door may come
back as \texttt{the corresponding resource is <url>} instead of the bare URL, an
Azure storage account URL may be split into an account name and a path that together
identify the same resource, or a bearer token may appear without its leading
\texttt{Bearer} keyword while the token itself is exactly right. In each case the
true value is present and unaltered, and a fair evaluation should credit it.

To handle these cases fairly without becoming permissive, we score predictions with
a three-step pipeline. Each step examines only the cases the previous one could not
resolve.

\paragraph{Step 1: Programmatic matching.} We first compare each prediction against
the ground truth automatically and collect every prediction that doesn't match
cleanly as a candidate false positive or false negative.

\paragraph{Step 2: LLM judge.} An evaluation agent reviews only those candidates
against explicit rules, reclassifying a prediction as correct when the true secret or
door is present in it despite surrounding text or reformatting, and leaving genuine
errors in place.

\paragraph{Step 3: Human review.} Because the judge is itself a language model and
the breadth of secret types produces edge cases the rules do not anticipate, a human
reviews any remaining candidate and makes the final decision case by case.

Only the labels that survive all three steps are used to compute the reported
precision and recall. The pipeline avoids two opposite failure modes: the spurious
errors produced by exact matching alone, and the unreliability of trusting an
automated judge without oversight.

\section{Results}
\label{sec:results}

\subsection{Multi-agent versus single-agent SSA}
\label{sec:results-mva}

We first examine whether the multi-agent design yields a meaningful improvement
over the single-agent variant. Both are evaluated on the multi-type benchmark of 356 synthetic emails, which contains 36 ground-truth secret instances across positive documents and mixes several secret--door types.Since large language models are
stochastic, we run every configuration five times and report the mean
together with a 95\% confidence interval. We evaluate four subtasks:
secret detection, door detection, secret extraction, and door extraction, under
precision and recall, and additionally report end-to-end latency across six
models of varying capability. Detection measures whether a secret or door is
identified at all, whereas extraction measures whether its exact value is
recovered.

For both variants, detection is close to saturated: precision remains above
87\%, and recall is near-perfect across all models. The verification agent
primarily improves precision. As \tableref{tab:secret} shows, secret-detection
precision rises from 87.48\% to 98.75\% on gpt-5.4-mini-medium and from 90.94\%
to 96.40\% on gpt-4.1-mini. This gain occasionally comes at a small cost in
recall; on gpt-5.4-mini-medium, secret-detection recall falls from 100\% to
97.33\%. Such behavior is expected when a critic is introduced, since filtering
candidate values to remove false positives can, in rare cases, also discard a
correct detection.

The clearest improvement appears in extraction, and particularly in door
extraction, as shown in \tableref{tab:door}. Multi-agent SSA raises
door-extraction precision for every model, with the largest gains on the weaker
models: precision improves by 16.3 points on gpt-4.1-mini, from 69.67\% to
86.00\%, and by 16.1 points on gpt-5.4-mini-medium, from 67.25\% to 83.38\%.
Secret-extraction precision follows the same pattern, again improving most on the
weakest model, gpt-4.1-mini, where it rises from 86.44\% to 95.26\%. As in
detection, this precision is obtained at a modest cost in recall: on gpt-4.1-mini
and gpt-5.4-mini-high, the single-agent variant retains slightly higher
door-extraction recall, because the review step occasionally removes a correct
door.

\begin{table}[t]
\floatconts
  {tab:secret}
  {\caption{Secret detection and extraction on the multi-type benchmark. Values
   are percentages, reported as mean $\pm$ 95\% confidence interval over five
   runs.}}
  {%
  \small
  \renewcommand{\arraystretch}{1.3}%
  \setlength{\tabcolsep}{5pt}%
  \begin{tabular}{@{}llcccc@{}}
  \toprule
  Model & Workflow & \shortstack{Detection\\Precision} & \shortstack{Detection\\Recall}
  & \shortstack{Extraction\\Precision} & \shortstack{Extraction\\Recall} \\
  \midrule
  \textbf{gpt-4.1-mini} & Single-agent & 90.94 $\pm$ 9.80 & \textbf{100.00 $\pm$ 0.00} & 86.44 $\pm$ 1.92 & 88.33 $\pm$ 2.89 \\
                        & Multi-agent  & \textbf{96.40 $\pm$ 6.59} & \textbf{100.00 $\pm$ 0.00} & \textbf{95.26 $\pm$ 8.64} & \textbf{91.67 $\pm$ 0.00} \\
  \midrule
  \textbf{gpt-5.4-mini-high} & Single-agent & \textbf{100.00 $\pm$ 0.00} & \textbf{100.00 $\pm$ 0.00} & 96.42 $\pm$ 3.10 & 89.44 $\pm$ 2.89 \\
                             & Multi-agent  & \textbf{100.00 $\pm$ 0.00} & 98.67 $\pm$ 3.70 & \textbf{100.00 $\pm$ 0.00} & \textbf{95.00 $\pm$ 5.67} \\
  \midrule
  \textbf{gpt-5.4-mini-medium} & Single-agent & 87.48 $\pm$ 6.62 & \textbf{100.00 $\pm$ 0.00} & 92.82 $\pm$ 2.02 & 86.11 $\pm$ 2.44 \\
                               & Multi-agent  & \textbf{98.75 $\pm$ 3.47} & 97.33 $\pm$ 4.53 & \textbf{98.46 $\pm$ 4.27} & \textbf{95.00 $\pm$ 5.67} \\
  \midrule
  \textbf{gpt-5.4-nano-high} & Single-agent & 95.00 $\pm$ 3.47 & \textbf{100.00 $\pm$ 0.00} & 94.49 $\pm$ 2.34 & 95.00 $\pm$ 1.54 \\
                             & Multi-agent  & \textbf{100.00 $\pm$ 0.00} & \textbf{100.00 $\pm$ 0.00} & \textbf{100.00 $\pm$ 0.00} & \textbf{96.67 $\pm$ 5.67} \\
  \midrule
  \textbf{gpt-5.4-nano-medium} & Single-agent & 93.75 $\pm$ 0.00 & \textbf{100.00 $\pm$ 0.00} & 94.46 $\pm$ 2.41 & 94.44 $\pm$ 2.44 \\
                               & Multi-agent  & \textbf{100.00 $\pm$ 0.00} & \textbf{100.00 $\pm$ 0.00} & \textbf{100.00 $\pm$ 0.00} & \textbf{98.33 $\pm$ 4.63} \\
  \midrule
  \textbf{glm-5.2} & Single-agent & 95.15 $\pm$ 6.17 & \textbf{100.00 $\pm$ 0.00} & 97.87 $\pm$ 2.74 & \textbf{100.00 $\pm$ 0.00} \\
                   & Multi-agent  & \textbf{100.00 $\pm$ 0.00} & \textbf{100.00 $\pm$ 0.00} & \textbf{100.00 $\pm$ 0.00} & \textbf{100.00 $\pm$ 0.00} \\
  \bottomrule
  \end{tabular}%
  }
\end{table}

\begin{table}[t]
\floatconts
  {tab:door}
  {\caption{Door detection and extraction on the multi-type benchmark. Values are
   percentages, reported as mean $\pm$ 95\% confidence interval over five runs.}}
  {%
  \small
  \renewcommand{\arraystretch}{1.3}%
  \setlength{\tabcolsep}{5pt}%
  \begin{tabular}{@{}llcccc@{}}
  \toprule
  Model & Workflow & \shortstack{Detection\\Precision} & \shortstack{Detection\\Recall}
  & \shortstack{Extraction\\Precision} & \shortstack{Extraction\\Recall} \\
  \midrule
  \textbf{gpt-4.1-mini} & Single-agent & 95.15 $\pm$ 6.17 & \textbf{100.00 $\pm$ 0.00} & 69.67 $\pm$ 3.55 & \textbf{95.76 $\pm$ 2.06} \\
                        & Multi-agent  & \textbf{96.40 $\pm$ 6.59} & \textbf{100.00 $\pm$ 0.00} & \textbf{86.00 $\pm$ 14.97} & 91.67 $\pm$ 0.00 \\
  \midrule
  \textbf{gpt-5.4-mini-high} & Single-agent & \textbf{100.00 $\pm$ 0.00} & \textbf{100.00 $\pm$ 0.00} & 77.37 $\pm$ 8.04 & \textbf{95.15 $\pm$ 2.06} \\
                             & Multi-agent  & \textbf{100.00 $\pm$ 0.00} & 98.67 $\pm$ 3.70 & \textbf{91.13 $\pm$ 8.88} & 95.00 $\pm$ 5.67 \\
  \midrule
  \textbf{gpt-5.4-mini-medium} & Single-agent & \textbf{100.00 $\pm$ 0.00} & \textbf{100.00 $\pm$ 0.00} & 67.25 $\pm$ 11.13 & 89.70 $\pm$ 7.80 \\
                               & Multi-agent  & 98.75 $\pm$ 3.47 & 97.33 $\pm$ 4.53 & \textbf{83.38 $\pm$ 11.12} & \textbf{95.00 $\pm$ 5.67} \\
  \midrule
  \textbf{gpt-5.4-nano-high} & Single-agent & 95.00 $\pm$ 3.47 & \textbf{100.00 $\pm$ 0.00} & 54.78 $\pm$ 3.54 & 70.30 $\pm$ 4.12 \\
                             & Multi-agent  & \textbf{100.00 $\pm$ 0.00} & \textbf{100.00 $\pm$ 0.00} & \textbf{66.81 $\pm$ 13.49} & \textbf{80.00 $\pm$ 13.88} \\
  \midrule
  \textbf{gpt-5.4-nano-medium} & Single-agent & 93.75 $\pm$ 0.00 & \textbf{100.00 $\pm$ 0.00} & 59.13 $\pm$ 11.50 & 72.73 $\pm$ 10.97 \\
                               & Multi-agent  & \textbf{100.00 $\pm$ 0.00} & \textbf{100.00 $\pm$ 0.00} & \textbf{68.36 $\pm$ 20.60} & \textbf{76.67 $\pm$ 18.51} \\
  \midrule
  \textbf{glm-5.2} & Single-agent & 95.15 $\pm$ 6.17 & \textbf{100.00 $\pm$ 0.00} & 97.87 $\pm$ 2.74 & \textbf{100.00 $\pm$ 0.00} \\
                   & Multi-agent  & \textbf{100.00 $\pm$ 0.00} & \textbf{100.00 $\pm$ 0.00} & \textbf{100.00 $\pm$ 0.00} & \textbf{100.00 $\pm$ 0.00} \\
  \bottomrule
  \end{tabular}%
  }
\end{table}

The per-model changes in \tableref{tab:secret-delta} and \tableref{tab:door-delta}
make this tradeoff more explicit. For secrets, the multi-agent workflow improves
extraction precision for all six models by 2.1 to 8.8 percentage points and
improves extraction recall by 0.0 to 8.9 points. Secret-detection precision also
increases for five of the six models, while recall is unchanged for four
models and decreases by at most 2.7 points. For doors, the precision gains are
larger: door-extraction precision increases by 2.1 to 16.3 points across all
models. Door-extraction recall is more mixed, ranging from a 4.1-point decrease
on gpt-4.1-mini to a 9.7-point increase on gpt-5.4-nano-high. Several of the
smaller detection changes, including the detection-recall decreases of at most
2.7 points, fall within the confidence intervals of the underlying measurements
and should be read as essentially unchanged rather than as reliable differences.
Overall, the verification agent mainly shifts SSA toward higher-precision
extraction, with the strongest benefit on the more ambiguous door field.

Two representative cases illustrate the mechanism behind these gains. In the
first, the detection agent recovered the secret but missed the associated
resource identifier, which the verification agent subsequently supplied,
improving door recall. In the second, the detection agent extracted an incorrect
resource URI, which the verification agent corrected, improving door precision.

These accuracy gains come at a cost. Because the multi-agent workflow issues an additional model call per document, it consumes more tokens and incurs higher latency and higher inference cost than the single-agent variant. As
\tableref{tab:latency} shows, the two-agent workflow runs roughly 1.2 to 2.3
times slower than the single-agent workflow. The largest relative slowdown occurs
on glm-5.2, where latency increases from 12142.8~ms to 28033.1~ms, or
2.31$\times$ slower; among the proprietary configurations the largest is
gpt-5.4-mini-medium, from 1417.5~ms to 2739.5~ms, or 1.93$\times$. The smallest
relative slowdown occurs on gpt-5.4-nano-medium, where latency increases from
2343.6~ms to 2712.6~ms, or 1.16$\times$ slower. This overhead makes multi-agent SSA better suited to asynchronous review, batch triage, or latency-tolerant workflows where the improved precision of the extracted secret--door pairs justifies the additional time and expense.

\begin{table}[t]
\floatconts
  {tab:latency}
  {\caption{End-to-end latency in milliseconds, averaged over five runs. Latency
   delta is the multi-agent latency divided by the single-agent latency. Bold
   marks the faster workflow per model.}}
  {%
  \small
  \renewcommand{\arraystretch}{1.3}%
  \setlength{\tabcolsep}{6pt}%
  \begin{tabular}{@{}lrrr@{}}
  \toprule
  Model & \shortstack{Single-agent\\latency} & \shortstack{Multi-agent\\latency} & \shortstack{Latency\\delta} \\
  \midrule
  \textbf{gpt-4.1-mini}        & \textbf{1611.0} & 2269.9 &  1.41$\times$ \\
  \textbf{gpt-5.4-mini-high}   & \textbf{2501.2}  & 4114.4 & 1.64$\times$ \\
  \textbf{gpt-5.4-mini-medium} & \textbf{1417.5}  & 2739.5 & 1.93$\times$ \\
  \textbf{gpt-5.4-nano-high}   & \textbf{2797.4}  & 3274.7 & 1.17$\times$ \\
  \textbf{gpt-5.4-nano-medium} & \textbf{2343.6}  & 2712.6 & 1.16$\times$ \\
  \textbf{glm-5.2}             & \textbf{12142.8} & 28033.1 & 2.31$\times$ \\
  \bottomrule
  \end{tabular}%
  }
\end{table}

The glm-5.2 results extend this picture to a larger, open-weight model from a
different family. Both variants are strong: the single-agent workflow already
reaches 100\% recall with 95.15\% detection and 97.87\% extraction precision, and
the review agent removes the residual false positives to bring every metric to
100\% (\tableref{tab:secret} and \tableref{tab:door}). This mirrors the stronger
proprietary configurations, where a capable base model leaves the reviewer little
to correct. That the same prompts and workflow perform well on a model from a different developer suggests that SSA is not tied to a single model family or vendor. The trade-off is speed: glm-5.2 is the slowest configuration we
evaluate and shows the steepest multi-agent slowdown (\tableref{tab:latency}),
so on this model the move to perfect precision is also the most expensive.

Finally, the benefit of the multi-agent design is model-dependent rather than
universal. The additional agent helps most when the single-agent baseline leaves more unsupported or incomplete extractions, where the review step can correct candidate errors and recover missing context. For stronger models, the
single-agent variant already performs comparably, and the added latency and cost
are less easily justified. We therefore regard multi-agent SSA as most valuable
when the underlying model requires additional reasoning support.

\subsection{Single-type versus multi-type secrets}
\label{sec:results-single-vs-multi}

To isolate the effect of credential diversity, we keep the method completely
fixed and vary only the data. We run the identical multi-agent SSA workflow on
both benchmarks, with the same detection-agent prompt and the same review-agent
prompt, the same six models, and the same five-run protocol used in
\sectionref{sec:results-mva}. The only difference is the benchmark: the
single-type benchmark of 204 emails restricts every secret to one type, the
Azure storage account key, whereas the multi-type benchmark mixes 23 secret
types in noisier documents. Any difference in performance between the two
therefore reflects the data alone, not a change in models, prompts, or
workflow.

Under this controlled setup, performance on the single-type benchmark is at or
near ceiling for every model. Detection precision is 100\% across all six models,
with recall at 100\% for the five proprietary configurations and 98.95\% for
glm-5.2, as shown in \tableref{tab:single-secret} and \tableref{tab:single-door}.
Extraction is almost as strong: secret-extraction precision is 100\% for every
model, with recall between 98.10\% and 99.52\%, and door extraction is perfect
for four of the six models, with gpt-5.4-nano-medium dipping to 98.00\% precision
and glm-5.2 to 98.95\% recall. The open-weight model is the informative case:
glm-5.2 comes from a different developer and a far larger size class than the mini
and nano configurations, yet it saturates the single-type benchmark just as they
do, its only shortfall being a single email missed at the detection stage in one
of five runs rather than any extraction error. Once the credential type is fixed,
the task is easy across model families and sizes alike, which is the baseline the
multi-type comparison departs from.

\begin{table}[t]
\floatconts
  {tab:single-secret}
  {\caption{Secret detection and extraction on the single-type benchmark with
   multi-agent SSA, where every secret is an Azure storage account key. Values are
   percentages, reported as mean $\pm$ 95\% confidence interval over five runs.}}
  {%
  \small
  \renewcommand{\arraystretch}{1.3}%
  \setlength{\tabcolsep}{5pt}%
  \begin{tabular}{@{}lcccc@{}}
  \toprule
  Model & \shortstack{Detection\\Precision} & \shortstack{Detection\\Recall}
  & \shortstack{Extraction\\Precision} & \shortstack{Extraction\\Recall} \\
  \midrule
  \textbf{gpt-4.1-mini}        & 100.00 $\pm$ 0.00 & 100.00 $\pm$ 0.00 & 100.00 $\pm$ 0.00 & 98.10 $\pm$ 1.32 \\
  \textbf{gpt-5.4-mini-high}   & 100.00 $\pm$ 0.00 & 100.00 $\pm$ 0.00 & 100.00 $\pm$ 0.00 & 99.05 $\pm$ 1.62 \\
  \textbf{gpt-5.4-mini-medium} & 100.00 $\pm$ 0.00 & 100.00 $\pm$ 0.00 & 100.00 $\pm$ 0.00 & 99.05 $\pm$ 1.62 \\
  \textbf{gpt-5.4-nano-high}   & 100.00 $\pm$ 0.00 & 100.00 $\pm$ 0.00 & 100.00 $\pm$ 0.00 & 99.05 $\pm$ 1.62 \\
  \textbf{gpt-5.4-nano-medium} & 100.00 $\pm$ 0.00 & 100.00 $\pm$ 0.00 & 100.00 $\pm$ 0.00 & 99.52 $\pm$ 1.32 \\
  \textbf{glm-5.2}             & 100.00 $\pm$ 0.00 & 98.95 $\pm$ 2.92 & 100.00 $\pm$ 0.00 & 98.95 $\pm$ 2.92 \\
  \bottomrule
  \end{tabular}%
  }
\end{table}

\begin{table}[t]
\floatconts
  {tab:single-door}
  {\caption{Door detection and extraction on the single-type benchmark with
   multi-agent SSA. Values are percentages, reported as mean $\pm$ 95\% confidence
   interval over five runs.}}
  {%
  \small
  \renewcommand{\arraystretch}{1.3}%
  \setlength{\tabcolsep}{5pt}%
  \begin{tabular}{@{}lcccc@{}}
  \toprule
  Model & \shortstack{Detection\\Precision} & \shortstack{Detection\\Recall}
  & \shortstack{Extraction\\Precision} & \shortstack{Extraction\\Recall} \\
  \midrule
  \textbf{gpt-4.1-mini}        & 100.00 $\pm$ 0.00 & 100.00 $\pm$ 0.00 & 100.00 $\pm$ 0.00 & 100.00 $\pm$ 0.00 \\
  \textbf{gpt-5.4-mini-high}   & 100.00 $\pm$ 0.00 & 100.00 $\pm$ 0.00 & 100.00 $\pm$ 0.00 & 100.00 $\pm$ 0.00 \\
  \textbf{gpt-5.4-mini-medium} & 100.00 $\pm$ 0.00 & 100.00 $\pm$ 0.00 & 100.00 $\pm$ 0.00 & 100.00 $\pm$ 0.00 \\
  \textbf{gpt-5.4-nano-high}   & 100.00 $\pm$ 0.00 & 100.00 $\pm$ 0.00 & 100.00 $\pm$ 0.00 & 100.00 $\pm$ 0.00 \\
  \textbf{gpt-5.4-nano-medium} & 100.00 $\pm$ 0.00 & 100.00 $\pm$ 0.00 & 98.00 $\pm$ 5.55 & 100.00 $\pm$ 0.00 \\
  \textbf{glm-5.2}             & 100.00 $\pm$ 0.00 & 98.95 $\pm$ 2.92 & 100.00 $\pm$ 0.00 & 98.95 $\pm$ 2.92 \\
  \bottomrule
  \end{tabular}%
  }
\end{table}

These results contrast sharply with the multi-type benchmark of
\sectionref{sec:results-mva}. The same workflow reaches 100\% door-extraction precision on the single-type benchmark for gpt-5.4-mini-medium and gpt-5.4-nano-high, and 98.00\% for gpt-5.4-nano-medium. On the multi-type benchmark, these scores fall to 83.38\%, 66.81\%, and 68.36\%, respectively, once secrets of 23 different types appear together, as reported in \tableref{tab:door}. Because the method is identical and only the benchmark changes, we attribute the multi-type degradation to properties of the data: greater credential diversity, noisier context, and adversarial constructions, rather than to a change in models, prompts, or workflow. SSA extracts a
known credential type almost perfectly; the multi-type setting is harder because
the model must additionally disambiguate among many types and resist distractor
and fragmented credentials.

\subsection{Human experts versus SSA}
\label{sec:results-human}
\FloatBarrier

Following the protocol in \sectionref{sec:methods-human-review}, we compare
multi-agent SSA against thirteen security analysts on the 45-document human-review
benchmark. Because both humans and SSA produce final secret--door pairs, we score
both at the pair level: a finding counts as correct only when both the secret and
its associated door match the ground truth. SSA detection precision and recall are
near ceiling across all models, so the comparison rests on extraction quality and
latency; the full per-model detection and extraction figures for SSA appear in
\appendixref{app:human-asf-full}.

\tableref{tab:human-vs-asf} places the two side by side. The volunteers reach
83.09\% precision but only 51.00\% recall, and on average a reviewer needs 53.12
minutes to work through all 45 documents, which comes to roughly 71 seconds per
document. Every SSA configuration is both more precise and far more complete, with
extraction precision above 95\% and recall above 95\%. The proprietary models
process each document in 4 to 14 seconds; the open-weight glm-5.2 is slower, at
roughly 40 seconds per document, while reaching the highest extraction precision.
\begin{table}[htbp]
\floatconts
  {tab:human-vs-asf}
  {\caption{Human security experts versus multi-agent SSA on the human-review
   benchmark, scored at the level of extracted secret--door pairs. Bold marks the best
   value in each column. Human values are reported as mean $\pm$ 95\% confidence interval across thirteen security experts; SSA values are reported as mean $\pm$ 95\% confidence interval over five runs. Full per-model detection and extraction figures for SSA
   appear in \appendixref{app:human-asf-full}.}}
  {%
  \small
  \renewcommand{\arraystretch}{1.3}%
  \setlength{\tabcolsep}{6pt}%
  \begin{tabular}{@{}lccc@{}}
  \toprule
  Method & \shortstack{Extraction\\Precision} & \shortstack{Extraction\\Recall}
  & \shortstack{Latency\\(s/doc)} \\
  \midrule
  Human experts & 83.09 $\pm$ 14.97 & 51.00 $\pm$ 15.97 & $\approx$71 \\
  \midrule
  SSA, gpt-4.1-mini        & 96.68 $\pm$ 1.43 & 97.90 $\pm$ 0.05 & \textbf{4.13} \\
  SSA, gpt-5.4-mini-high   & 98.41 $\pm$ 1.11 & 99.59 $\pm$ 1.13 & 13.93 \\
  SSA, gpt-5.4-mini-medium & 97.75 $\pm$ 0.92 & \textbf{99.64 $\pm$ 1.01} & 6.99 \\
  SSA, gpt-5.4-nano-high   & 95.87 $\pm$ 4.79 & 97.02 $\pm$ 3.01 & 12.86 \\
  SSA, gpt-5.4-nano-medium & 97.13 $\pm$ 2.88 & 98.32 $\pm$ 1.17 & 11.00 \\
  SSA, glm-5.2             & \textbf{100.00 $\pm$ 0.00} & 95.13 $\pm$ 5.03 & 40.47 \\
  \bottomrule
  \end{tabular}%
  }
\end{table}
\FloatBarrier

The recall gap is the most consequential difference. Working by hand at the same volume, the human reviewers miss roughly half of the exposed secret--door pairs across 45 multi-format documents, whereas SSA maintains recall above 97\% across all evaluated configurations. For incident response, where a single undetected secret can leave a
door open, this difference in completeness, delivered in seconds rather than nearly
an hour, is precisely the gap that automation is meant to close.

\subsection{SSA versus traditional secret scanners}
\label{sec:results-baselines}
\FloatBarrier
We next compare SSA against two widely used secret scanners, TruffleHog and
Nightfall, on the multi-type secret dataset. These tools detect secret strings but
do not infer or pair the associated door, so we compare at the level of secret
detection only, scored over the 36 secret instances in the dataset. We averaged
Nightfall and SSA over five runs and report mean and 95\% confidence interval.
TruffleHog detects secrets with fixed regular-expression patterns, so its output is
deterministic and identical on every run; repeated trials add no information, and
we therefore ran it once and report a single value without an interval.
\tableref{tab:baselines} reports the results.

\begin{table}[ht]
\floatconts
  {tab:baselines}
  {\caption{Secret detection on the multi-type dataset: SSA versus Nightfall and TruffleHog. Values are percentages. Nightfall and SSA are averaged over five runs
   (mean $\pm$ 95\% CI). 
   $^{\dagger}$TruffleHog is deterministic and was run once, so has no CI.}}
  {%
  \small
  \renewcommand{\arraystretch}{1.3}%
  \setlength{\tabcolsep}{6pt}%
  \begin{tabular}{@{}lcc@{}}
  \toprule
  Method & \shortstack{Detection\\Precision} & \shortstack{Detection\\Recall} \\
  \midrule
  Nightfall   & 74.07 $\pm$ 0.00 & 55.56 $\pm$ 0.00 \\
  TruffleHog$^{\dagger}$ & \textbf{100.00} & 27.78 \\
  \midrule
  SSA, gpt-4.1-mini        & 96.40 $\pm$ 6.59 & \textbf{100.00 $\pm$ 0.00} \\
  SSA, gpt-5.4-mini-high   & \textbf{100.00 $\pm$ 0.00} & 98.67 $\pm$ 3.70 \\
  SSA, gpt-5.4-mini-medium & 98.75 $\pm$ 3.47 & 97.33 $\pm$ 4.53 \\
  SSA, gpt-5.4-nano-high   & \textbf{100.00 $\pm$ 0.00} & \textbf{100.00 $\pm$ 0.00} \\
  SSA, gpt-5.4-nano-medium & \textbf{100.00 $\pm$ 0.00} & \textbf{100.00 $\pm$ 0.00} \\
    SSA, glm-5.2             & \textbf{100.00 $\pm$ 0.00} & \textbf{100.00 $\pm$ 0.00} \\
  \bottomrule
  \end{tabular}%
  }
\end{table}

The two scanners fall short of SSA in opposite ways. Nightfall finds 20 of the 36
secret instances but adds seven false positives per run, leaving it at 74.07\%
precision and 55.56\% recall. TruffleHog raises no false positives and so reaches
100\% precision, but it recovers only 10 of the 36 instances, a recall of 27.78\%.
SSA avoids both failure modes. Recall is 100\% for four of the six models and
never falls below 97.33\%, and precision is at or above 96.40\% throughout,
reaching 100\% for four configurations. Four of the six SSA configurations match
TruffleHog's perfect precision while more than tripling its recall, and every
configuration exceeds Nightfall on precision and recall at once.

Neither baseline reports doors. TruffleHog and Nightfall return the secret string
and stop there, so a responder still has to track down what each credential
unlocks. SSA returns the secret, its door, and the supporting evidence together,
which is the result a responder can act on.

\FloatBarrier
\section{Discussion}
\label{sec:discussion}

Taken together, our studies support the central claim behind SSA: an agentic
large-language-model workflow can recover exposed credentials from unstructured
documents and identify the resource each credential may unlock, accurately enough
to support first-pass triage in our synthetic benchmarks.

\paragraph{From detecting secrets to naming what they unlock.} Our framing of secret finding as secret--door extraction separates SSA from the scanners we evaluate. The baseline comparison in \sectionref{sec:results-baselines} shows the
gap concretely: In our benchmark, TruffleHog and Nightfall detect secret strings but do not identify the associated door, leaving responders to determine what resource the credential may unlock. SSA instead returns the secret, its candidate door, and supporting evidence in one result. The detection numbers alone understate this
difference, because even when a scanner finds the secret it never produces the pair
that tells a responder where to act. Recovering the door is what helps turn a detection into an actionable triage finding.

\paragraph{A first-pass triage layer.} In practice, SSA is best viewed as a
first-pass triage layer rather than a replacement for existing scanners or analyst
judgment. A responder or product workflow can run SSA over exposed documents such as
emails, tickets, chat transcripts, and incident notes, and receive structured
findings containing the secret, candidate door, and supporting evidence.
These findings can be routed to downstream validation, credential rotation,
ownership lookup, or remediation workflows. Analysts stay in the loop for
prioritization and final judgment, while SSA reduces the manual effort of
discovering what leaked and where responders should act.

\paragraph{When the review agent earns its cost.} The multi-agent design pays off
unevenly, and \sectionref{sec:results-mva} shows where. Adding a review agent mainly raises precision, with the largest gains on door extraction and on configurations where the single-agent baseline leaves more unsupported or incomplete outputs. Door-extraction precision improves by as much as 16 points. The cost
is latency: the second pass runs the workflow 1.2 to 2.3 times slower. This points
to a clear deployment rule. Multi-agent SSA is most worthwhile when the single-agent baseline leaves enough errors to justify a second pass, or when review is asynchronous and latency is less critical. The single-agent variant is a reasonable choice when a strong model already extracts cleanly and latency matters. The review step is a lever for precision, not a universal default.

\paragraph{Credential diversity is the hard part.} Holding the method fixed and
varying only the data isolates what actually makes extraction difficult. In
\sectionref{sec:results-single-vs-multi}, the same workflow that reaches
near-perfect precision and recall on a single credential type drops sharply once 23
types appear together in noisier documents. The difficulty appears to come from the combined effects of credential diversity, noisier context, and adversarial constructions in the multi-type benchmark, rather than from the Azure storage account key task alone. Our synthetic pipeline and evaluation
framework made this controlled comparison possible, and the practical lesson is that
single-type results can overstate readiness: a system should also be measured on mixed,
distractor-laden documents that resemble real exposed content.

\paragraph{Augmenting security analysts, not replacing them.} The comparison in
\sectionref{sec:results-human} argues for redirecting analyst effort, not removing
it. Scanning long documents for every embedded secret and door is tedious, detail-oriented work, and our human-review benchmark shows that findings can slip through even with experienced reviewers. By
handling this first pass in seconds with high recall, SSA frees analysts for the
judgments that need security expertise, such as assessing what an exposed credential
puts at risk, prioritizing remediation, and validating whether a credential is still
live.

\paragraph{Future work.} The clearest next step is validation on real, sanctioned incident data, which would test how well performance on the synthetic benchmark transfers to operational settings.
Knowing the door also opens a path to active verification: once SSA identifies the resource a credential may unlock, the system could route the finding to provider-supported or policy-approved validity checks and tell a responder not only what leaked, but whether the credential appears to remain active. Reducing the cost
of the review agent, for example by routing only uncertain or weak-model cases
through a second pass, would make multi-agent SSA practical in latency-sensitive
settings. We also plan to broaden the document formats, credential types, and languages, and to study how SSA fits into responder workflows as a first-pass triage tool that supports human judgment.

\acks{We thank Dr.\ Malachi Jones for his guidance and for providing valuable
feedback throughout the development of this paper.

\noindent We are also grateful to the thirteen security experts who volunteered their time to
take part in our human-agent review study: Noah Baertsch, Haley Bui-Nguyen, Cory
Clowes, Hemal Desai, Siva Gangadhar Galla, Stanley He, Blaine Herro, Joe Mansour,
Manuel Mel\'endez, Cristal Ruiz, Mauricio Velazco, Isabella White, and Limin Yang. Their
willingness to carefully read every
document and record each secret--door pair, without tooling or AI assistance, made the
human comparison in this work possible. We especially thank them for sharing the
operational security expertise they have built over years of hands-on incident-response
work; their judgment grounded our evaluation in real-world practice and meaningfully
strengthened this paper.

\noindent\textbf{Disclaimer.} This paper describes a research prototype evaluated on
synthetic data. It is not a product announcement, availability commitment, or roadmap
statement. All metrics reflect performance on the synthetic benchmarks described herein
and are not guarantees of detection or security outcomes. Third-party product names are
used for factual comparison only and remain the property of their respective owners; no
partnership or endorsement is implied.}

\appendix

\clearpage
\section{Worked example: secret, door, and evidence}
\label{app:example}

The following synthetic example illustrates how SSA turns an unstructured
request into a secret, its associated door, and the supporting evidence that
links them. The secret shown is a non-functional placeholder.

\begin{center}
\fbox{\begin{minipage}{0.92\linewidth}
\textbf{Input document snippet}\par\medskip
``Can someone rotate the key for the staging storage account? The account is
\texttt{contosostaging}, and the blob endpoint is
\url{https://contosostaging.blob.core.windows.net/}. The key currently used by
the migration job is \texttt{8f3Hk2pQ9rLmEXAMPLEKEYa1b2c3d4e5f6==}. Please
update the pipeline after rotation.''

\medskip
\textbf{SSA output}\par\medskip
\textbf{Secret:} \texttt{8f3Hk2pQ9rLmEXAMPLEKEYa1b2c3d4e5f6==}\par
\textbf{Door:} \url{https://contosostaging.blob.core.windows.net/}\par
\textbf{Evidence:} The document links the key to the \texttt{contosostaging}
storage account and its blob endpoint.\par
\textbf{Actionability:} A responder can identify both the leaked credential and
the storage resource that may need validation, rotation, or access review.
\end{minipage}}
\end{center}

\renewcommand{\thetable}{\thesection\arabic{table}}
\renewcommand{\thefigure}{\thesection\arabic{figure}}

\clearpage
\section{Types of secrets}
\label{app:secrets-doors}
\setcounter{table}{0}\setcounter{figure}{0}

The multi-type benchmark spans the 23 common secret types listed in
\tableref{tab:secret-types}. Each type is paired with a corresponding door,
the resource it unlocks.

\begin{table}[H]
\floatconts
  {tab:secret-types}
  {\caption{The 23 secret types in the multi-type benchmark. Each type is paired
   with a corresponding door, the resource it unlocks.}}
  {%
  \small
  \renewcommand{\arraystretch}{1.15}%
  \setlength{\tabcolsep}{6pt}%
  \begin{tabular}{@{}l p{0.55\linewidth}@{}}
  \toprule
  Secret type & Description \\
  \midrule
  \texttt{azure\_storage\_key}        & Azure Storage account key \\
  \texttt{azure\_storage\_sas}        & Azure Storage shared access signature (SAS) token \\
  \texttt{aws\_access\_key\_id}       & AWS access key ID \\
  \texttt{aws\_secret\_access\_key}   & AWS secret access key \\
  \texttt{aws\_session\_token}        & AWS temporary session token \\
  \texttt{aws\_s3\_presigned\_url}    & AWS S3 pre-signed URL \\
  \texttt{gcp\_service\_account\_key} & GCP service account key \\
  \texttt{gcp\_signed\_url}           & GCP signed URL \\
  \texttt{entra\_password}            & User or application password \\
  \texttt{entra\_aad\_client\_secret} & Entra (AAD) application client secret \\
  \texttt{entra\_aad\_cert\_thumbprint} & Entra (AAD) certificate thumbprint \\
  \texttt{entra\_aad\_certificate\_pem} & Entra (AAD) certificate (PEM) \\
  \texttt{oauth\_bearer\_token}       & OAuth bearer token (\texttt{Authorization: Bearer ...}) \\
  \texttt{api\_key}                   & Generic API key, e.g.\ \texttt{sk\_live\_...}, \texttt{api\_key=...}, \texttt{x-api-key:...} \\
  \texttt{service\_token}             & Service-to-service token \\
  \texttt{jwt\_token}                 & JSON Web Token (three dot-separated base64url parts) \\
  \texttt{session\_cookie}            & Session cookie \\
  \texttt{basic\_auth\_header}        & HTTP Basic auth header (\texttt{Authorization: Basic base64(user:pass)}) \\
  \texttt{magic\_login\_link}         & Email login, reset, or magic link that grants access \\
  \texttt{db\_username\_password}     & Database username and password \\
  \texttt{db\_connection\_string}     & Database connection string \\
  \texttt{ssh\_private\_key\_pem}     & SSH/PEM private key (\texttt{-----BEGIN ... PRIVATE KEY-----}) \\
  \texttt{generic\_secret}            & Any secret enabling sensitive access not covered above \\
  \bottomrule
  \end{tabular}%
  }
\end{table}

\section{Per-model changes from single-agent to multi-agent SSA}
\label{app:deltas}
\setcounter{table}{0}\setcounter{figure}{0}

\tableref{tab:secret-delta} and \tableref{tab:door-delta} report the per-model
percentage-point change from single-agent to multi-agent SSA on the multi-type
benchmark, derived from \tableref{tab:secret} and \tableref{tab:door}. Positive
values indicate an improvement under the multi-agent workflow.

\begin{table}[t]
\floatconts
  {tab:secret-delta}
  {\caption{Change in secret detection and extraction from single-agent SSA to
   multi-agent SSA. Values are percentage-point changes.}}
  {%
  \small
  \renewcommand{\arraystretch}{1.3}%
  \setlength{\tabcolsep}{6pt}%
  \begin{tabular}{@{}lrrrr@{}}
  \toprule
  Model & \shortstack{Detection\\precision} & \shortstack{Detection\\recall}
  & \shortstack{Extraction\\precision} & \shortstack{Extraction\\recall} \\
  \midrule
  \textbf{gpt-4.1-mini}        & +5.5  &  0.0 & +8.8 & +3.3 \\
  \textbf{gpt-5.4-mini-high}   &  0.0  & -1.3 & +3.6 & +5.6 \\
  \textbf{gpt-5.4-mini-medium} & +11.3 & -2.7 & +5.6 & +8.9 \\
  \textbf{gpt-5.4-nano-high}   & +5.0  &  0.0 & +5.5 & +1.7 \\
  \textbf{gpt-5.4-nano-medium} & +6.3  &  0.0 & +5.5 & +3.9 \\
  \textbf{glm-5.2}             & +4.9  &  0.0 & +2.1 &  0.0 \\
  \bottomrule
  \end{tabular}%
  }
\end{table}

\begin{table}[H]
\floatconts
  {tab:door-delta}
  {\caption{Change in door detection and extraction from single-agent SSA to
   multi-agent SSA. Values are percentage-point changes.}}
  {%
  \small
  \renewcommand{\arraystretch}{1.3}%
  \setlength{\tabcolsep}{6pt}%
  \begin{tabular}{@{}lrrrr@{}}
  \toprule
  Model & \shortstack{Detection\\precision} & \shortstack{Detection\\recall}
  & \shortstack{Extraction\\precision} & \shortstack{Extraction\\recall} \\
  \midrule
  \textbf{gpt-4.1-mini}        & +1.3 &  0.0 & +16.3 & -4.1 \\
  \textbf{gpt-5.4-mini-high}   &  0.0 & -1.3 & +13.8 & -0.2 \\
  \textbf{gpt-5.4-mini-medium} & -1.3 & -2.7 & +16.1 & +5.3 \\
  \textbf{gpt-5.4-nano-high}   & +5.0 &  0.0 & +12.0 & +9.7 \\
  \textbf{gpt-5.4-nano-medium} & +6.3 &  0.0 & +9.2  & +3.9 \\
  \textbf{glm-5.2}             & +4.9 &  0.0 & +2.1  &  0.0 \\
  \bottomrule
  \end{tabular}%
  }
\end{table}

\section{Full SSA detection and extraction on the human-review benchmark}
\label{app:human-asf-full}
\setcounter{table}{0}\setcounter{figure}{0}

\begin{table}[H]
\floatconts
  {tab:human-secret}
  {\caption{Detection and extraction by multi-agent SSA on the human-review
   benchmark. Values are percentages, reported as mean $\pm$ 95\% confidence
   interval over five runs.}}
  {%
  \small
  \renewcommand{\arraystretch}{1.3}%
  \setlength{\tabcolsep}{5pt}%
  \begin{tabular}{@{}lcccc@{}}
  \toprule
  Model & \shortstack{Detection\\Precision} & \shortstack{Detection\\Recall}
  & \shortstack{Extraction\\Precision} & \shortstack{Extraction\\Recall} \\
  \midrule
  \textbf{gpt-4.1-mini}        & 100.00 $\pm$ 0.00 & 100.00 $\pm$ 0.00 & 96.68 $\pm$ 1.43 & 97.90 $\pm$ 0.05 \\
  \textbf{gpt-5.4-mini-high}   & 100.00 $\pm$ 0.00 & 100.00 $\pm$ 0.00 & 98.41 $\pm$ 1.11 & 99.59 $\pm$ 1.13 \\
  \textbf{gpt-5.4-mini-medium} & 100.00 $\pm$ 0.00 & 100.00 $\pm$ 0.00 & 97.75 $\pm$ 0.92 & 99.64 $\pm$ 1.01 \\
  \textbf{gpt-5.4-nano-high}   & 100.00 $\pm$ 0.00 & 98.26 $\pm$ 2.96 & 95.87 $\pm$ 4.79 & 97.02 $\pm$ 3.01 \\
  \textbf{gpt-5.4-nano-medium} & 100.00 $\pm$ 0.00 & 100.00 $\pm$ 0.00 & 97.13 $\pm$ 2.88 & 98.32 $\pm$ 1.17 \\
  \bottomrule
  \end{tabular}%
  }
\end{table}

\bibliography{new_bib}

\end{document}